\documentclass[aip,jap,letterpaper,amsmath,amssymb,reprint]{revtex4-1}
\usepackage{graphicx}
\usepackage{microtype}
\usepackage[utf8]{inputenc} 
\usepackage{float}

\begin{document}
\title{Improved thermal stability in doped MnN/CoFe exchange bias systems}
\author{M. Dunz}
 \email{mdunz@physik.uni-bielefeld.de}
\author{B. Büker}%
\author{M. Meinert}%
\affiliation{ 
$^1$Center for Spinelectronic Materials and Devices, Department of Physics, Bielefeld University, D-33501 Bielefeld, Germany 
}%
\date{\today}

\begin{abstract}
We investigated the influence of doping antiferromagnetic MnN in polycrystalline MnN/CoFe exchange bias systems, showing high exchange bias of up to 1800\,Oe at room 
temperature. The thermal stability of those systems is limited by nitrogen diffusion that occurs during annealing processes. In order to improve the thermal stability, defect energies of elements throughout the periodic table substituting Mn were calculated via density functional theory. Elements calculated to have negative defect energies bind nitrogen stronger to the lattice and could be able to prevent diffusion. We prepared exchange bias stacks with doping concentrations of a few percent by (reactive) co-sputtering, testing doping elements with defect energies ranging from highly negative to slightly positive. We show that doping with elements calculated to have negative defect energies indeed improves the thermal stability. Y doped MnN layers with doping concentrations below 2\% result in systems that show exchange bias fields higher than 1000\,Oe for annealing temperatures up to $485\,^{\circ}$C.
\end{abstract}

\maketitle

\begin{figure*}[t!]
\centering
\includegraphics[width=13cm]{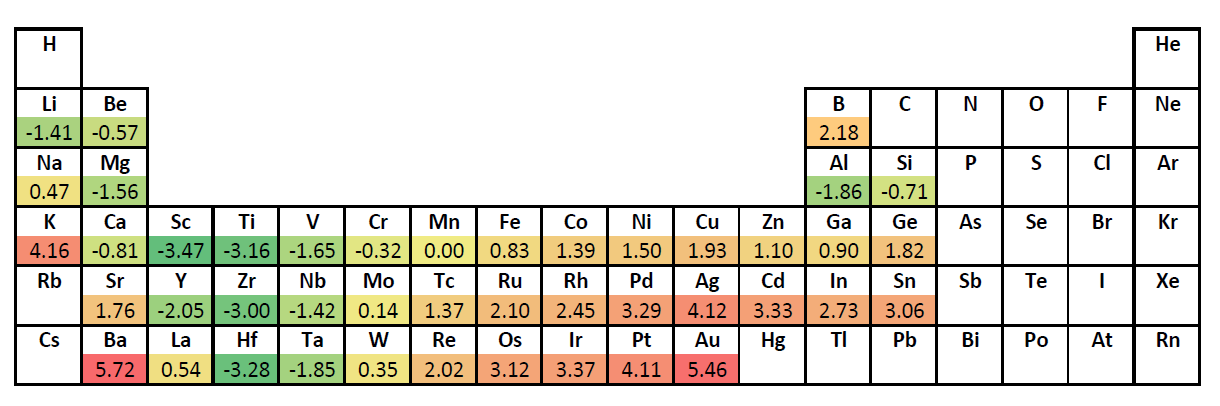}
\caption{Defect energies (eV) for replacing an Mn atom in the MnN lattice calculated
with density functional theory for different elements of the periodic table.}
\label{energies}
\end{figure*}
\begin{figure}[t!]
\centering
\includegraphics[width=9cm]{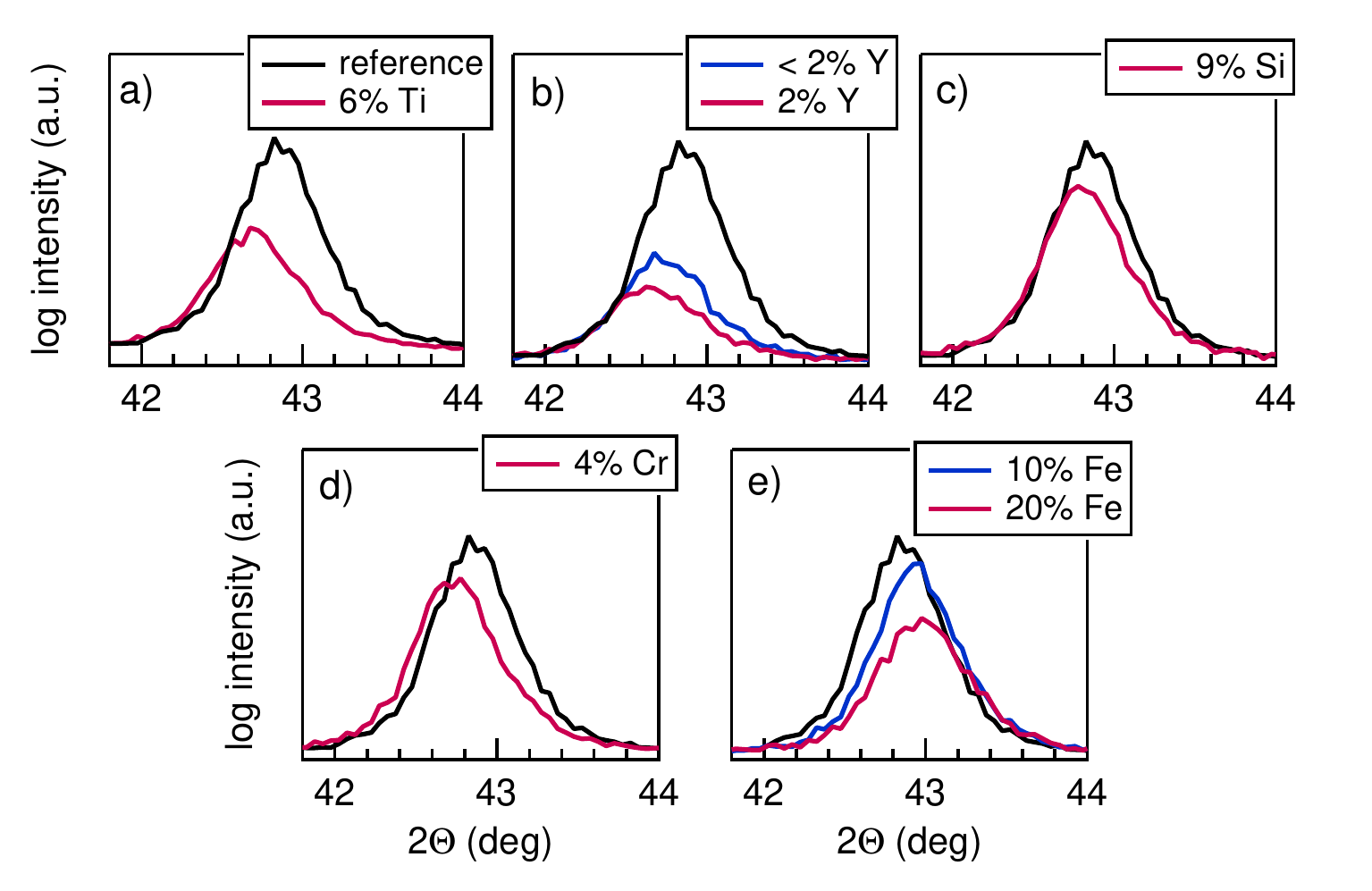}
\caption{X-ray diffraction spectra around the angular range of the (002) peak of MnN for exchange bias stacks with differently doped MnN layers with $t_{\text{MnN+dopant}}=30$\,nm: a) Ti, b) Y, c) Si, d) Cr, e) Fe. In each graph, a reference peak of undoped MnN is shown. All presented samples were annealed at $325\,^{\circ}$C prior to the XRD measurements. The graph's order from a) to e) relates to increasing defect energy of the dopants, i.e. less strong binding of nitrogen.}
\label{dopedXRD}
\end{figure}
\begin{figure}[h!]
\centering
\includegraphics[width=8.8cm]{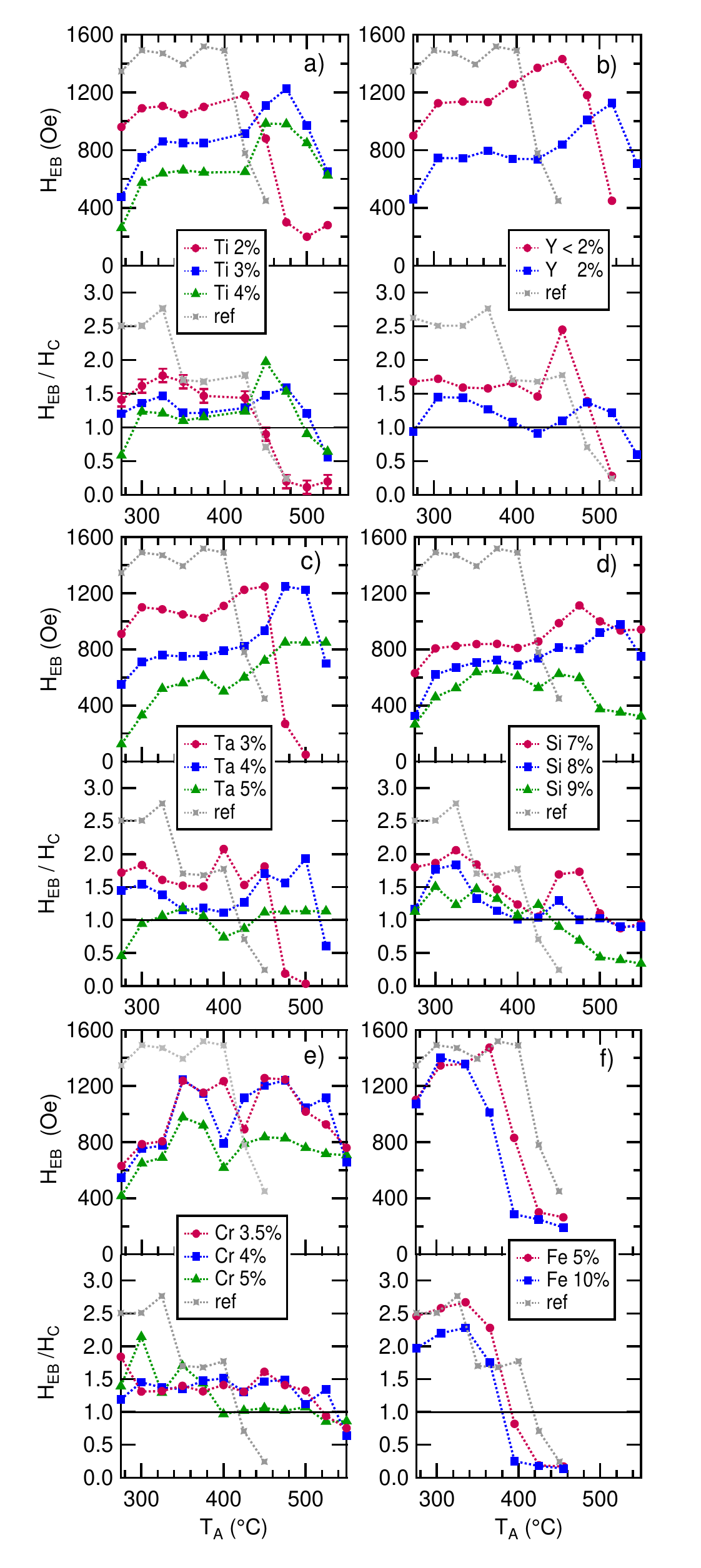}
\caption{Annealing series detected for doped stacks with $t_{\text{MnN+dopant}}=30$\,nm. Exchange bias and ratio $H_{\text{EB}}/H_{\text{C}}$ are shown in dependence on the annealing temperature for samples where MnN was doped with different concentrations of a) Ti, b) Y, c) Ta, d) Si, e) Cr and f) Fe. A series measured on an undoped MnN/CoFe system is shown for reference (ref). The error bars are smaller than the markers.}
\label{series}
\end{figure}

In spinelectronics, the exchange bias effect\citep{Meiklejohn, Schuller, Berkowitz, Kiwi, Grady} is used to pin a ferromagnetic electrode to an antiferromagnetic layer. This is crucial in GMR or TMR stacks to allow for distinct stable resistance states\citep{Chappert}. For several years, the search for new antiferromagnetic materials for exchange bias has been going on in order to find rare-earth free alternatives for commonly used MnIr\citep{Funke} or MnPt\citep{Saito, Glaister}. For integration into spinelectronic devices, the antiferromagnet should be easy to prepare, generate exchange bias fields that are clearly higher than corresponding coercive fields and be thermally stable at typical device operation temperatures. As we recently reported\citep{Meinert2015, Zilske, Dunz}, antiferromagnetic MnN is a very promising candidate. \\
MnN crystallizes in the $\Theta-$phase of the Mn-N phase diagram\citep{Gokcen}, a tetragonal variant of the NaCl structure with $a = b = 4.256$\,\AA\, and $c = 4.189$\,\AA\, at room temperature\citep{Suzuki2000}. The exact lattice constants depend on the nitrogen content in the lattice. With increasing nitrogen content, increasing lattice constants are observed\citep{Suzuki2000, Leineweber}. Optimized polycrystalline MnN/CoFe bilayer systems show exchange bias of up to 1800 Oe at room temperature with an effective interfacial exchange energy of $J_\mathrm{eff} = 0.41$\,mJ/m$^2$ and an effective uniaxial anisotropy constant of $K_\mathrm{eff} =  37$\,kJ/m$^3$\,\citep{Meinert2015}. They yield ratios of $H_{\text{eb}}/H_{\text{c}}$ significantly larger than one and are easy to prepare with sputter deposition at room temperature, satisfying earlier mentioned requirements for integration into spintronic devices. The Néel temperature of MnN is around $660$\,K\citep{Tabuchi} and MnN/CoFe systems show a median blocking temperature of $160\,^{\circ}$C\citep{Meinert2015}. However, nitrogen diffusion at high temperatures, respectively long annealing times, limits the thermal stability of the system. In the course of our previous investigations\citep{Meinert2015} we already found that preparing MnN with a higher nitrogen concentration can slightly increase the thermal stability but at the same time lowers the exchange bias.\\
In the present article, we show that doping MnN with small concentrations of elements that strengthen the nitrogen bonds also improves its thermal stability. To find suitable dopants, defect energies of elements substituting Mn were calculated with density functional theory (DFT) using the Vienna Ab Initio Simulation Package (VASP)\citep{vasp}. The calculations were performed using supercells consisting of 16 unit cells with one Mn atom replaced. The calculated defect formation energies are similar to the formation enthalpies of the respective nitrides. The results are depicted in Figure \ref{energies}. Elements exhibiting negative defect energies ensure that the nitrogen atoms are bonded stronger to the lattice compared to undoped MnN, whereas elements with positive defect energies are unlikely to mix.\\
To investigate the influence of different dopants on MnN, we prepared film stacks of Ta 10\,nm / (doped) MnN 30\,nm / Co$_{70}$Fe$_{30}$ 1.6\,nm / Ta 0.5\,nm / Ta$_2$O$_5$ 2\,nm on thermally oxidized SiO$_{\text{x}}$ substrates via magnetron (co-)sputtering at room temperature. Undoped samples were prepared for reference. We followed exactly the same preparation procedure as described in our previous report\citep{Meinert2015}. The base pressure of the sputtering system was around $5 \times 10^{-9}$\,mbar prior to the deposition runs. The MnN films were reactively sputtered from an elemental Mn target with a gas ratio of  $50\%$ Ar to $50\%$ N$_2$ at a working pressure of $p_{\text{w}}=2.3\cdot10^{-3}$\,mbar. The typical deposition rate of MnN was 0.1\,nm/s at a source power of 50\,W. The doped MnN films were co-sputtered using the same sputter parameters for MnN as for the undoped samples. The doping element was sputtered from an RF source at low power to achieve small doping concentrations. Approximate elemental ratios were calculated from the specific deposition rates, densities and molar masses prior to the deposition to estimate the respective doping concentrations. As we worked with very small amounts of the doping material, a reliable exact determination of the doping concentration via X-ray fluoresence was not possible. Subsequent post-annealing for 15 min and field cooling in a magnetic field of $H_{\text{fc}}=6.5$\,kOe parallel to the film plane was performed in a vacuum furnace with pressure below $5\cdot 10^{-6}$\,mbar to activate exchange bias. Magnetic characterization of the samples was performed using the longitudinal magneto-optical Kerr effect (MOKE) at room temperature. For annealing series, samples were successively annealed and measurements were taken in between the single steps. Structural analysis was performed via X-ray diffraction with a Philips X'Pert Pro MPD, which is equipped with a Cu source and Bragg-Brentano optics.\\
\begin{figure}[b!]
\centering
\includegraphics[width=9cm]{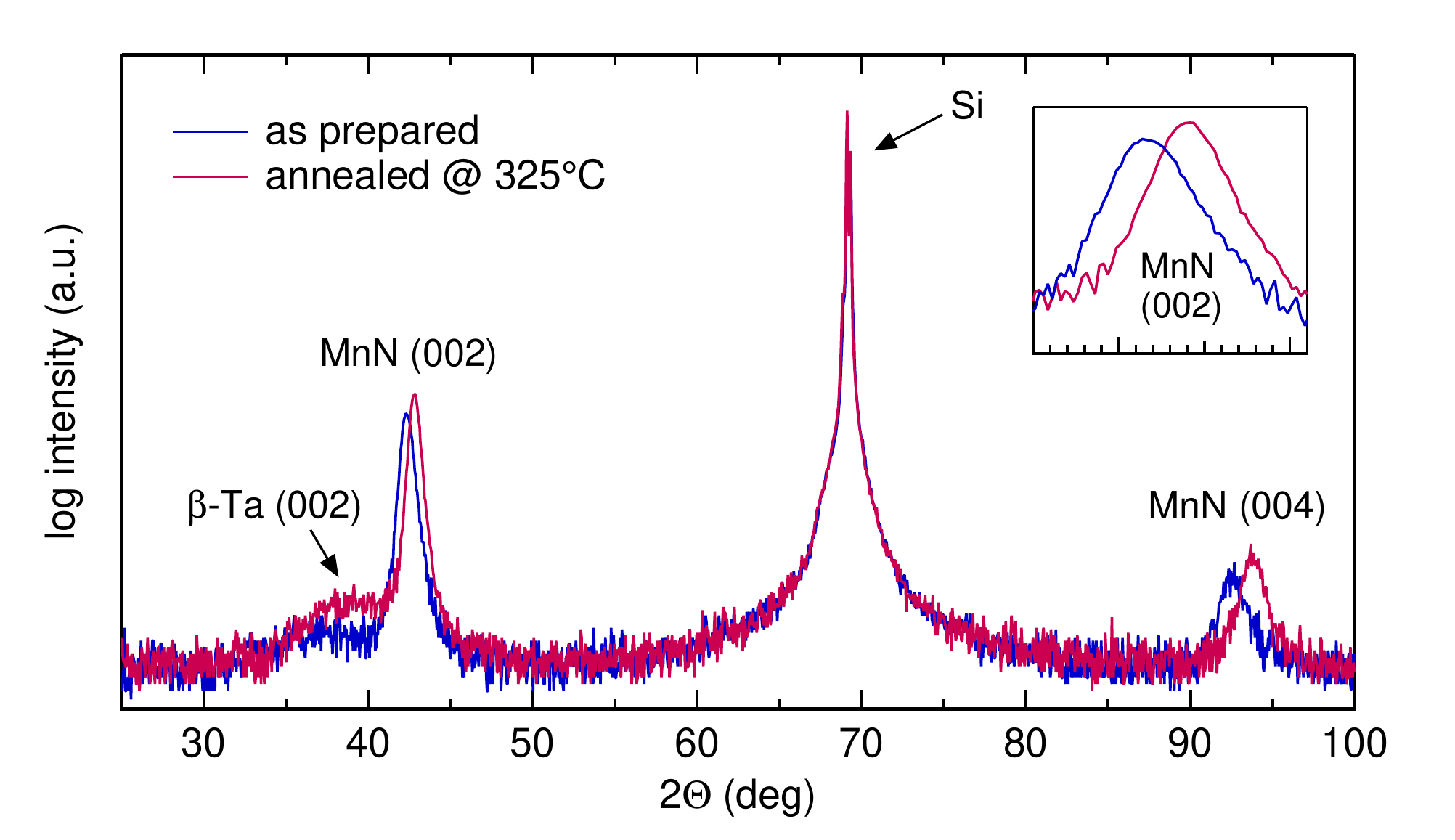}
\caption{X-ray diffraction pattern of an undoped exchange bias stack with $t_{\text{MnN}}=30$\,nm. The blue curve displays the spectrum measured directly after preparation; the red one after annealing the sample at $325\,^{\circ}$C.}
\label{xray}
\end{figure}
To verify the growth of the $\Theta-$phase of MnN, X-ray diffraction measurements of samples with undoped MnN films were performed initially. In Fig. \ref{xray}, a typical diffraction spectrum of the Ta/MnN/CoFe stacks before and after annealing at $325\,^{\circ}$C is shown. All peaks are identified as belonging to the substrate, to the $\beta$-Ta seed layer, or to the MnN film. The lattice constant of MnN directly after preparation is $c = 4.272$\,\AA, which is slightly larger than the bulk values reported in the literature\citep{Suzuki2000}. The measurements revealed a polycrystalline, columnar growth of the MnN in (001) direction. After annealing, the MnN peaks are slightly shifted to higher angles, indicating a smaller lattice parameter of $c = 4.216$\,\AA. This can be attributed to the generation of nitrogen vacancies as interdiffusion occurs during the annealing process, leading to a decrease of the MnN's lattice constants.\\
To investigate the influence of doping on the crystal structure of MnN, additional XRD scans were taken for the doped samples. Figure \ref{dopedXRD} shows diffraction scans around the angular range of the (002) peak of MnN for differently doped MnN films compared to an undoped reference. All samples were annealed at $325\,^{\circ}$C prior to the measurement. The graph's order from a) to e) relates to increasing defect energy of the dopants, i.e. presumably weaker binding of nitrogen. The (002) peak of MnN layers doped with elements calculated to have a negative defect energy, namely Ti, Y, Si or Cr, is located at smaller angles with respect to the reference, indicating larger lattice constants. This hints at the fact that in the doped samples less nitrogen diffuses out of the MnN layer during annealing as a result of the enhanced nitrogen bonds. Vice versa, doping with an element that has a positive defect energy, like Fe, results in a shift of the (002) peak to higher angles compared to the reference. Here, stronger diffusion occurs during the annealing process because of the weaker binding of nitrogen. It has to be pointed out that we do not know to which extent the  different atomic radii of the dopants might have an influence when it comes to modified lattice constants. Dopants with larger atomic radii than Mn could increase the lattice constants, overlaying the effect of reduced nitrogen diffusion. However, even though Ti, Y and Ta do indeed have a larger atomic radius than Mn\citep{atomradii}, the other dopants have an atomic radius that is either comparable (Cr, Fe) or smaller (Si) than the one of Mn\citep{atomradii} and yet altered lattice constants can be measured. We thus conclude that this results from a different nitrogen content in these systems compared to undoped MnN, caused by stronger or weaker binding of nitrogen. \\
For the magnetic characterization, detailed annealing series were performed in order to
test the influence of the different dopants on the thermal stability of the MnN/CoFe system in terms of exchange bias. Figure \ref{series} displays the dependence of exchange bias on the annealing temperature for samples where MnN was doped with Ti, Y, Ta, Si, Cr, or Fe. Additionally, the ratio of exchange bias and coercive field $H_{\text{EB}}/H_{\text{C}}$ is depicted. For each doping material, samples with different doping concentrations were probed. Independent of the element, we found that the resulting exchange bias decreases with increasing doping concentration. This behavior is probably attributed to modifications in the crystal or the magnetic structure when too much doping material is incorporated. This is in line with the decreased crystallinity that can be observed in Figure \ref{dopedXRD} for the elements Y and Fe when going to higher doping concentrations. In general, the expectation that doping elements calculated to have negative defect energies enhance the thermal stability of the CoFe/MnN system is fulfilled: despite smaller maximum values, exchange bias can be observed for annealing temperatures higher than $500\,^{\circ}$C for systems doped with Ti, Y , Ta , Si  or Cr. The undoped reference sample loses its exchange bias after annealing at $425\,^{\circ}$C.\\
With $-3.16$\,eV, Ti is calculated to have the most negative defect energy of all dopants, i.e. the lowest formation energy. Figure \ref{series}a) shows that exchange bias is still observable after annealing at $525\,^{\circ}$C in systems where MnN was doped with $3\%$ or $4\%$ Ti. Doping with $2\%$ Ti leads to higher exchange bias values but does not increase the thermal stability significantly. Concerning the ratio $H_{\text{EB}}/H_{\text{C}}$, doping with a concentration of $3\%$ Ti yields the best results, showing values higher than one for annealing temperatures up to $500\,^{\circ}$C. \\
Doping MnN with Y, calculated to have a defect energy of $-2.05$\,eV, also significantly increases the thermal stability of our exchange bias system as can be seen in Figure \ref{series}b): a doping concentration of only $2\%$ Y results in an increase of thermal stabiliy of $150\,^{\circ}$C in terms of exchange bias. However, $H_{\text{EB}}/H_{\text{C}}>1$ is not fulfilled throughout the whole annealing temperature range. Going to a smaller doping concentration of Y yields stable exchange bias up to annealing temperatures of $500\,^{\circ}$C with values that are clearly higher than the corresponding coercive fields. Due to the very low deposition rate of Y for this small doping concentration, the resulting elemental ratio could not be determined clearly. As the Y source power was set to a slightly smaller value than for the preparation of the sample doped with  $2\%$ Y, we can only estimate that the doping concentration is also smaller.\\
Ta is calculated to have a slightly less negative defect energy of $-1.85$\,eV. Figure \ref{series}c) shows that again doping with a high concentration of $5\%$ Ta leads to stable exchange bias up to annealing temperatures above $500\,^{\circ}$C but can not guarantee $H_{\text{EB}}/H_{\text{C}}>1$ due to lower total values. Smaller doping concentrations of $3\%$ and $4\%$ Ta increase the thermal stability of the system to $450\,^{\circ}$C and $500\,^{\circ}$C, respectively, and yield ratios  $H_{\text{EB}}/H_{\text{C}}$ that are clearly higher than one after annealing up to those temperatures. \\
In Figure \ref{series}d), the results for doping MnN with Si, calculated defect energy of $-0.71$\,eV, are displayed. Due to a very high sputter rate, no doping concentrations of less than $7\%$ Si could be obtained with the standard preparation parameters. Doping with these high concentrations of $7\%$, $8\%$ and $9\%$ Si results in very low exchange bias values that are not significantly higher than the corresponding coercive fields throughout the whole annealing temperature range. However, an increase of thermal stability in terms of exchange bias is clearly oberservable. For systems doped with $8\%$ and $9\%$ Si, exchange bias fields around $800$\,Oe are measured after annealing at $550\,^{\circ}$C. \\
With $-0.32$\,eV, Cr is calculated to have the least negative defect energy of all elements we investigated. Nevertheless, it seems particularly interesting as a doping element due to its magnetic character. The DFT calculations predict that the Cr atoms exhibit a magnetic moment of $2.5$\,$\mu_{\text{B}}$ that is comparable to the one of the Mn atoms with $2.8$\,$\mu_{\text{B}}$. Figure \ref{series}e) shows that doping with Cr also increases the thermal stability but yields very fluctuating exchange bias values with a distinct minimum between $400\,^{\circ}$C and $450\,^{\circ}$C for all doping concentrations. Other striking differences to the doping materials with vanishing magnetic moments are not observable. Again, the two smaller concentrations of $3.5\%$ and $4\%$ Cr yield higher absolute values and a ratio $H_{\text{EB}}/H_{\text{C}}>1$ up to $500\,^{\circ}$C, whereas doping with $5\%$ Cr causes the ratio to fall below one after annealing at much lower temperatures. \\
In contrast to all previously investigated doping elements, Fe is calculated to have a positive defect energy of $0.83$\,eV, weakening the nitrogen bonds. Hence, doping MnN with Fe is expected to reduce its thermal stability. Our results shown in Figure \ref{series}f) can clearly confirm this assumption. Doping MnN with a concentration of $5\%$ Fe leads to stable exchange bias values up to $365\,^{\circ}$C whereas a doping concentration of $10\%$ Fe already yields declining exchange bias after annealing at temperatures higher than $335\,^{\circ}$C. The thermal stability is thus reduced by nearly $70\,^{\circ}$C. However, the exchange bias as well as the ratio $H_{\text{EB}}/H_{\text{C}}$ of the doped samples show values that are comparable to the undoped reference for lower annealing temperatures, especially when doping with $5\%$ Fe. Such high values could not be observed for any doping material with a negative defect energy.\\
In summary, we prepared polycrystalline exchange bias systems with doped MnN/CoFe bilayers that show exchange bias fields of more than 1000~Oe and ratios $H_{\text{EB}}/H_{\text{C}}>1$. Doping MnN influences the thermal stability of the system in dependence on the dopant's ability to bind the nitrogen. Essentially, we investigated doping with elements calculated to have negative defect energies and thus strengthen the nitrogen bonds, thereby preventing diffusion during the annealing processes. We tested the influence of Ti, Ta, Y, Cr, and Si on MnN and found that all elements indeed lead to an enhanced thermal stability in terms of stable exchange bias after annealing at high temperatures. However, the improved stability comes with lower total exchange bias values for higher doping concentrations. The best results were obtained when doping MnN with a very small amount of less than $2\%$ Y, yielding high exchange bias of clearly larger than 1000~Oe for annealing temperatures up to $485\,^{\circ}$C. This relates to an increased thermal stability of nearly $100\,^{\circ}$C. Additionally, we tried doping MnN with Fe, calculated to have a positive defect energy weakening the bonds, and consistently found that the resulting exchange bias system was thermally less stable than the undoped reference. Our experiments thus confirm the results of the previously performed DFT calculations.

\end{document}